\documentclass[aps, letter, twocolumn,superscriptaddress]{revtex4-2}

\usepackage[english]{babel}
\usepackage{amssymb,amsmath}
\usepackage{graphicx}
\usepackage{float}
\usepackage{placeins}

\begin{document}
	
	\title{Exposing impostor Majorana zero modes through atomic-scale shot-noise}
	
	\author{A. Maiti}
	\email[]{abhishek.cnrs@gmail.com}
	\affiliation{Laboratoire de Physique des Solides (CNRS UMR 8502), B\^{a}timent 510, Universit\'{e} Paris-Saclay, 91405 Orsay, France}
	
	\author{G. D. Gu}
	\affiliation{Condensed Matter Physics and Materials Science Department, Brookhaven National Laboratory, Upton, NY, USA}
	
	\author{F. Massee}
	\email[]{freek.massee@universite-paris-saclay.fr}
	\affiliation{Laboratoire de Physique des Solides (CNRS UMR 8502), B\^{a}timent 510, Universit\'{e} Paris-Saclay, 91405 Orsay, France}
	
	\date{\today}
	
	\begin{abstract}
		A robust zero-bias conductance peak in putative $p$-wave superconductors is often regarded as the primary signature of a Majorana zero mode. Yet similar features can also arise from trivial bound states. This ambiguity has limited the reliability of conventional spectroscopy as a diagnostic tool, raising a long-standing problem of how to detect such impostors. Here, we address this issue with an alternative approach, atomic-scale shot-noise spectroscopy, that goes beyond conductance measurements. Through a detailed investigation of multiple defect-bound zero-bias states in the widely studied superconductor Fe(Se,Te), we observe that differential conductance can exhibit an apparently `robust' zero-bias peak. However, shot-noise measurements consistently reveal the fingerprint of the individual particle- and hole character hidden in the tunnelling conductance, unambiguously exposing the trivial nature of the zero-bias peak. Our results establish shot-noise spectroscopy as a decisive diagnostic for ruling out false Majorana signatures in atomic-scale experiments.
		
	\end{abstract}
	
	\maketitle
	
	\noindent\textbf{\textit{Introduction.---}} The search for non-Abelian states of matter has become a central theme of modern quantum materials research \cite{Nayak_RMP}, motivated by the prospect of Majorana bound states as building blocks for topological qubits \cite{Kitaev_AnnPhys}. Experimental efforts to realise such states have largely focused on engineered effective $p$-wave superconductivity in hybrid platforms, including semiconductor nanowires \cite{Lutchyn_NRM}, magnetic atom chains \cite{NadjPerge_PRB2013, Pientka_PRB2013, NadjPerge_Science2014, Kim_SciAdv2018, Schneider_Nano2022}, topological insulators \cite{Xu_PRL2015_1}, and two-dimensional ferromagnets \cite{PalacioMorales_SciAdv2019, Kezilebieke_Nature2020}, as well as bound states in iron-based superconductors \cite{Xu_PRL2016, Wang_PRB2015, Zhang_Science2018, Zhang_NPhys2019, Yin_NPhys2015, Chen_NPhys2020, Fan_NC2021, Liu_PRX2018, Wang_Science2018, Machida_NMat2019, Zhu_Science2020, Liu_NC2020, Kong_NC2021}.
	In this direction, scanning tunnelling microscopy (STM), with its atomic-scale spatial and high energy resolution, has emerged as a key probe \cite{Jack_NRevPhys2021, rachel_pr_2025}, with zero-bias conductance peaks widely regarded as signatures of Majorana modes \cite{Xu_PRL2015_1, Yin_NPhys2015, Liu_PRX2018, Wang_Science2018, Machida_NMat2019, Chen_NPhys2020, Zhu_Science2020, Liu_NC2020, Kong_NC2021, Fan_NC2021, Pientka_PRB2013, NadjPerge_Science2014, Menard_NC2017, Kim_SciAdv2018, PalacioMorales_SciAdv2019, Kezilebieke_Nature2020, Schneider_Nano2022}. 
	
	Similar zero-energy features can also arise from topologically trivial states, including Yu-Shiba-Rusinov (YSR) states \cite{Heinrich_PSS, Ruby_PRL2015_1} and overlapping Caroli–de Gennes–Matricon vortex states \cite{Kim_APR2021, Ge_NC2023}, necessitating detailed spectroscopic tests under varying perturbations \cite{Yin_NPhys2015, Chatzopoulos_NC2021, Uldemolins_NC2024, Zhang_NanoLett, Lee_NanoLett2025}. As an additional diagnostic, spin-polarised STM has emerged as a promising technique \cite{Sun_PRL2016, Wang_PRL2021}, but requires a magnetic tip, an external magnetic field and, to extract the spin signature, not a single measurement, but a difference between two measurements, adding a layer of complexity. 
	
	The idea to, instead, go beyond the conductance and examine the tunnelling dynamics by shot-noise measurements was proposed early on in several theoretical studies \cite{Bolech2007, Nilsson2008, Cao2012, Liu2015, Perrin_PRB2021, Wong2022, mei_prb_2024}. Although most proposals are experimentally challenging to implement, a minimal single-tip tunnelling framework \cite{Perrin_PRB2021}, compatible with existing setups \cite{Massee_RSI, Bastiaans_RSI}, predicts that the shot-noise-derived effective Fano factor $F^*$ \cite{Ya_PhysRep2000} can serve as a decisive diagnostic. This is because while trivial in-gap states show inelastic quasiparticle relaxation and Andreev processes, yielding $F^* \neq 1$ \cite{Thupakula_PRL2022, Ohnmacht_PRR}, Majorana-assisted perfect Andreev reflection uniquely gives $F^* = 1$, providing a quantitative criterion to distinguish true Majorana modes from near-zero-energy YSR states (see Supplemental Material \cite{SI}, Sec I). The key open question is if the deviations from $F^* = 1$ observed on finite energy YSR states ($E_{\mathrm{YSR}} > k_B T$) that are in excellent agreement with the theoretical model \cite{Thupakula_PRL2022}, can still experimentally be observed for states at, or near, zero-bias ($E_{\mathrm{YSR}} < k_B T$), which mimic a Majorana zero mode.
	
	Near zero-bias, electron-hole thermal mixing affects the conductance asymmetry, and may similarly affect shot-noise, potentially leading to $F^*=1$ even for trivial states. Indeed, observations of $F^* = 1$ for both trivial zero-energy vortex states and candidate Majorana vortices by a previous noise study \cite{Ge_NC2023} suggest this possibility. However, the same study also found no clear distinction between the tunnelling dynamics in finite-energy YSR states and the bare superconducting substrate, possibly due to the unconventional measurement protocol that is used. This, in combination with the limited resolution of the study and the absence of a theoretical framework for experiments with a superconducting tip, makes interpretation of these results inconclusive. In this letter we present a proof-of-principle experimental study showing that, unlike previous work, a single shot-noise measurement can indeed reveal the trivial YSR nature of a (near-) zero-bias peak and thereby provide a direct diagnostic to identify impostor Majoranas.
	
\begin{figure}[h]
		\centering
		\includegraphics[width=0.75\textwidth]{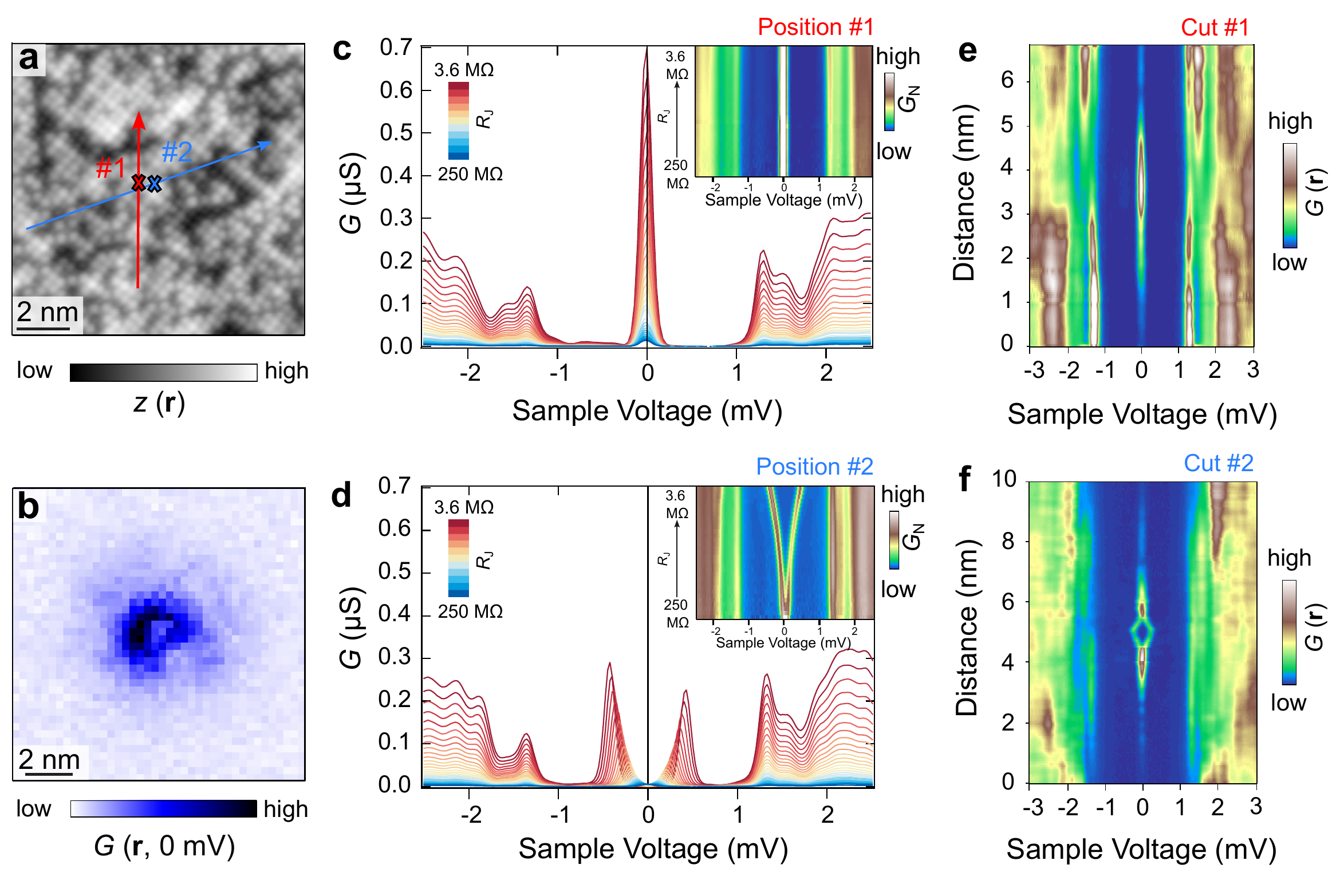}
		\caption{
			\textbf{Trivial YSR states mimicking Majorana-like zero-bias peak.}
			\textbf{a} High resolution topography ($V_{\text{bias}} = 50$ mV, $I = 1.5$ nA) of Fe(Se,Te), showing a clean surface with Se (bright) and Te (dark) contrast. 
			\textbf{b} Zero-bias conductance map on the same location as panel a, revealing an in-gap state induced by a sub-surface impurity (defect \#A). Set conditions: $V_{\text{bias}} = 10$ mV, $I = 100$ pA.
			\textbf{c,d} Differential conductance measured at position \#1 and position \#2, respectively for different junction resistances ($R_J$) as indicated in the legend. The step size in resistance uses a logarithmic scale. Inset shows the 2D map of the spectra (normalised by the setup current).
			\textbf{e,f} Differential conductance taken along the cut through position \#1 and position \#2, respectively marked by the red and blue arrows in \textbf{b}.}
		\label{fig:1}
	\end{figure}

	\noindent\textbf{\textit{Experimental Results.---}} To address whether a trivial YSR state (near) zero-energy still shows its characteristic signature in the noise, we turn our attention to defect-induced states in superconducting Fe(Se,Te). Fig.~\ref{fig:1}a shows a topography of the Fe(Se,Te) surface, with the distinctive signature of the Se and Te atoms on the surface. Despite the absence of visible surface defects,
	\clearpage
	\onecolumngrid
	
	\twocolumngrid
	
	\noindent a map of the zero-bias differential conductance ($G(\mathbf{r}, E) = dI/dV$) (Fig.~\ref{fig:1}b) reveals a localised spectroscopic feature, similar to those observed in prior works \cite{Chatzopoulos_NC2021, Wang_Science2020}. To demonstrate that conductance spectroscopy alone could lead to false Majorana interpretations, we examine this sub-surface impurity (defect \#A) in more detail. In particular, we expose the defect to different tip-induced electric fields, which has been shown in previous work to be able to shift or split sub-gap states in Fe(Se,Te) \cite{Chatzopoulos_NC2021, Uldemolins_NC2024, Zhang_NanoLett}. At high junction resistances ($>$100~M$\Omega$), the defect shows a sharp zero-bias resonance regardless of the tip location, which remains at zero for lower junction resistances on the majority of the defect wave function (see Fig.~\ref{fig:1}c). However, directly above the impurity (position \#2), the zero-bias peak splits into two non-zero peaks (Fig.~\ref{fig:1}d) for low junction resistances, revealing the trivial nature of the defect state. Here, the effect is much less pronounced than for impurities on the surface \cite{Uldemolins_NC2024}, and in case of the defect in Fig.~\ref{fig:1}a splitting is only observed exactly on top of the burried impurity. To highlight the highly localised nature of the effect, and to show how field induced splitting can easily be overlooked, we took linecuts along two trajectories. Whereas the cut in Fig.~\ref{fig:1}f reveals the splitting, Fig.~\ref{fig:1}e shows no splitting along the length of the cut, masking its trivial nature. 
	
	In numerous previous STM studies, signatures similar to those observed at position \#1  were interpreted as Majorana zero modes at defects \cite{Yin_NPhys2015, Chen_NPhys2020, Fan_NC2021}, vortices \cite {Liu_PRX2018, Wang_Science2018, Machida_NMat2019, Zhu_Science2020, Liu_NC2020, Kong_NC2021}, at the boundary of a two-dimensional superconductor \cite{Menard_NC2017}, in ferromagnet-superconductor hetero-structures \cite{PalacioMorales_SciAdv2019, Kezilebieke_Nature2020}, and at edge of atomic chains on superconductors \cite {Pientka_PRB2013, NadjPerge_Science2014, Kim_SciAdv2018, Schneider_Nano2022}. Here, we clearly show that conductance spectroscopy can - in absence of an extensive experimental survey (see Supplemental Material \cite{SI}, Sec III for details) - in principle lead to a false Majorana interpretation. Importantly, not all systems respond to a changing electric field between the tip and sample as strongly as Fe(Se,Te) does, necessitating more complex conductance measurements with spin-polarised tips \cite{Sun_PRL2016, Wang_PRL2021} and/or quantised conductance measurements \cite{Zhu_Science2020} to expose impostor Majoranas. We thus proceed to address the central question: can shot-noise serve as a more reliable single shot diagnostic to detect the trivial nature of (near-) zero-bias states?
	
	To investigate the tunnelling process into the defect-bound zero-bias peak, we measured shot-noise by employing our home-built cryogenic circuitry, as described elsewhere \cite{Massee_RSI}. To calibrate our circuitry, we first measured noise in a clean region of the sample, i.e. away from defects such that conductance is fully gapped (see Supplemental Material \cite{SI}, Sec II, IV and V). The effective Fano factor extracted from our fit is $F^* = 1$ outside the gap, consistent with elastic quasi-particle tunnelling with a Poissonian statistics \cite{Ronen_PNAS2016}.
	
\begin{figure}[htbp]
		\centering
		\includegraphics[width=1\columnwidth]{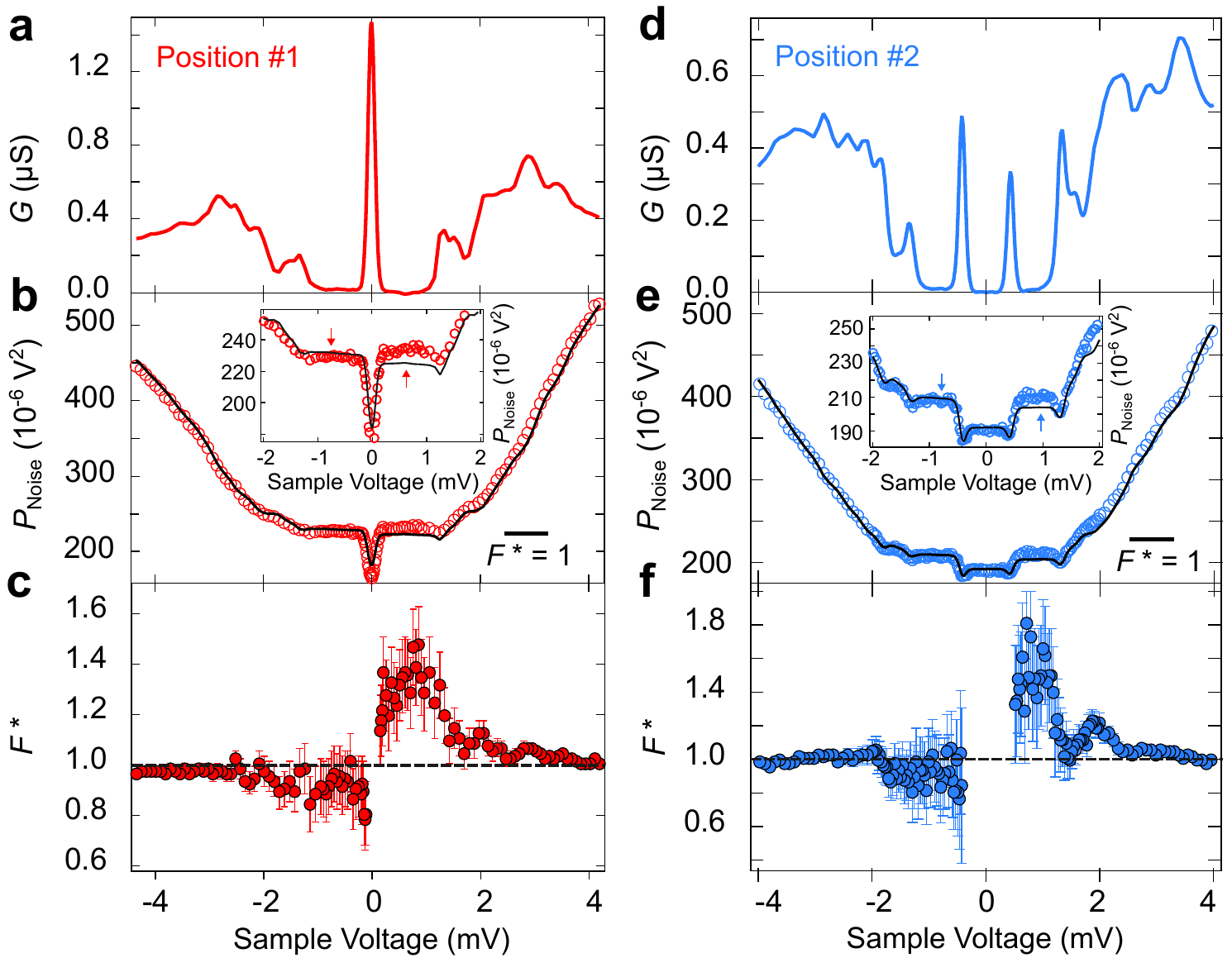}
		\caption{
			\textbf{Shot-noise spectroscopy of the zero-bias and split peak(s).}
			\textbf{a} Differential conductance measured at position ~\#1 (see Fig.~\ref{fig:1}a). Set-point conditions: $V_{\text{bias}} = 4.8~\mathrm{mV}$, $I = 1.5~\mathrm{nA}$.  
			\textbf{b} Simultaneously recorded current noise. The inset highlights the region of interest around the zero-bias peak. The black line indicates the reference $F^* = 1$. Deviations from this line are observed on both the positive and negative flanks of the zero-bias peak, marked by the red arrows.  
			\textbf{c} Extracted effective Fano factor, $F^*$, as a function of sample voltage, clearly showing $F^*\neq1$ in the region of interest.
			\textbf{d-f} A similar data set, using the same set-point conditions at position~\#2, also shows $F^* \neq 1$ at the split YSR peaks.}
		\label{fig:2}
	\end{figure}

	In Fig.~\ref{fig:2}a, we present the voltage-dependent noise at position \#1 showing the zero-bias conductance peak. Since the current goes to zero at zero-bias, determining $F^*$ exactly at zero-bias is experimentally impossible. However, the data taken slightly away from zero-bias, where the current is still dominated by the zero-bias peak, for example at $\pm$ 0.5~mV, can reliably probe the tunnelling dynamics of the zero-bias peak. This is because as long as the current is (near-) exclusively carried by the zero bias state, the noise is also solely reflecting the zero-bias state. Using all circuit parameters as extracted from the reference measurement, as well as the measured dynamical resistance (see Supplemental Material \cite{SI}, Sec II, IV), the only free parameter is the effective Fano factor $F^*$. Clearly, a fit assuming $F^* = 1$, which is expected for a true Majorana due to a perfect Andreev process, fails to describe the measured noise inside the gap (Fig.~\ref{fig:2}b).
	
	To show this more clearly, we extract the $F^*(V)$ curve by allowing only $F^*$ to vary for each voltage in the fit. This result (Fig.~\ref{fig:2}c) demonstrates that at $+$0.5~mV the noise is significantly enhanced ($F^* > 1$), by approximately $\sim 40 \%$, whereas, at $-$0.5~mV, the noise is slightly reduced ($F^* < 1$) by roughly $\sim 10 \%$. Current dependent measurements at fixed voltage (Supplemental Material \cite{SI}, Sec VI) confirm the robustness of our findings and rule out spurious sources of noise. Although at the negative polarity the deviation in absolute numbers is relatively small, the enhancement at the positive polarity is much larger than the experimental error (Supplemental Material \cite{SI}, Sec IV).
	
	We carried out a control measurement at position~\#2 under same conditions. The results (Fig.~\ref{fig:2}d-f) for the split peak are near-identical: outside the gap, $F^* = 1$, while inside the gap at the two YSR peaks, the noise is either enhanced ($F^* > 1$) or reduced ($F^* < 1$), depending on the bias polarity. The behaviour of noise enhancement and suppression naturally follows from the electron ($u$)-hole ($v$) asymmetry of a pair of YSR peaks, $G_{\pm}$ = $G(\pm E_\text{YSR})$, where $E_\text{YSR}$ is the YSR resonance energy ($G_{+}\propto u^{2}$, $G_{-}\propto v^{2}$). The smaller peak primarily reflects coherent Andreev processes ($F^* > 1$), while the larger peak is dominated by single-electron tunnelling and inelastic relaxation ($F^* < 1$). This is because the larger resonance necessarily acquires an inelastic contribution, because the smaller resonance cannot support Andreev processes for all carriers contributing to the larger peak. In contrast, the inelastic contribution is diminished when tunnelling occurs into the smaller resonance, particularly at low $R_J$. 
	
	The simultaneous observation of noise enhancement and reduction at the zero-bias peak confirms that the zero-bias peak is actually an overlapping pair of trivial YSR states whose energy separation, $2E_\text{YSR}$, is smaller than the thermal broadening, $\Delta E = 3.5\,k_{\mathrm{B}}T \approx 0.15~\mathrm{mV}$ for $T = 0.5~\mathrm{K}$. Although the trivial nature cannot be detected at position~\#1 solely from conductance spectroscopy, shot-noise immediately reveals it in a single measurement by showing $F^* \neq 1$.
	
\begin{figure}[htbp]
		\centering
		\includegraphics[width=1\columnwidth]{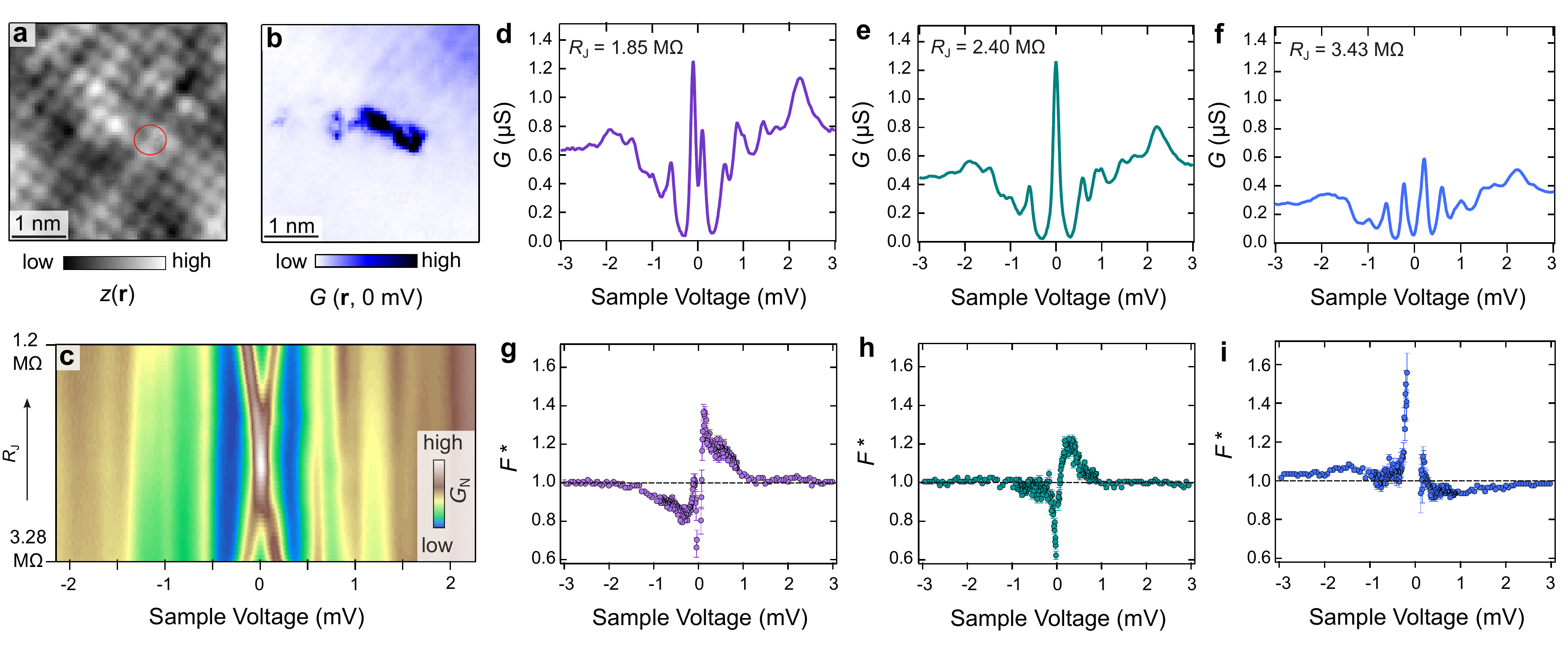}
		\caption{
			\textbf{Shot-noise spectroscopy revealing the trivial nature across a quantum phase transition.}
			\textbf{a} simultaneously taken topography and zero-bias conductance map of the sub-surface impurity (defect \#B) showing a QPT. Set-point conditions: $V_{\text{bias}} = 4.4$ mV, $I = 1$ nA). The red circle denotes the area where the spectroscopic measurements were carried out.
			\textbf{c} Junction resistance ($R_J$) dependence of the differential conductance (normalised by the set-point current) measured inside the impurity ranging from $1.2~\mathrm{M\Omega}$ (top) to $3.28~\mathrm{M\Omega}$ (bottom). The step size in $R_J$ uses a logarithmic scale.
			\textbf{d-f} Differential conductance measured for three different resistances: $1.85~\mathrm{M\Omega}$ ($G_{-} > G_{+}$), $2.4~\mathrm{M\Omega}$ (crossing), and $3.43~\mathrm{M\Omega}$ ($G_{-} < G_{+}$). The $R_J$ values required to access different regimes of the QPT are spatially sensitive, although this does not affect the general idea of the measurement. 
			\textbf{g-i} Effective Fano factor, $F^*$, as a function of sample voltage. See Supplemental Material \cite{SI}, Sec VII, for the voltage dependent noise data. The electron-hole (a)symmetry is directly reflected in the reduction and enhancement of noise from unity. Note that the asymmetry in $F^*$ is also flipped consistently with the relative wright of $G_{-}, G_{+}$ in \textbf{d,f}. Crucially, for the zero-bias peak (\textbf{e,h}), a similar reduction and enhancement of noise ($F^* \neq 1$) indicates that it is also a pair of YSR peaks near the crossing.}
		\label{fig:3}
	\end{figure}

	To reinforce the findings of our study, we present a different impurity (defect \#B) in Fig.~\ref{fig:3}a,b, where, by changing $R_J$ while keeping the xy-position fixed, we can tune the energy of the in-gap state through zero-bias. This change signals a quantum phase transition (QPT) from a screened (singlet) to an unscreened (doublet) impurity ground state (Fig.~\ref{fig:3}c) \cite{Heinrich_PSS}. Importantly, the state crossing zero is well isolated from other states allowing us to measure the shot-noise in three characteristic regimes of the QPT: on either side of the QPT, where $G_{-} > G_{+}$ (Fig.~\ref{fig:3}d) and $G_{-} < G_{+}$ (Fig.~\ref{fig:3}f), and near/at the crossing point where $G_{+}$ and $G_{-}$ merge into a single resonance (Fig.~\ref{fig:3}e). 
	The $F^*(V)$ curves (Fig.~\ref{fig:3}g-i) also track the phase transition. In the regime $G_{-} > G_{+}$, we find $F^* < 1$ on the negative-energy side and $F^* > 1$ on the positive-energy side. When the asymmetry is reversed, $G_{-} < G_{+}$, the asymmetry in $F^*$ is likewise inverted, with $F^* < 1$ on the positive-energy side and $F^* > 1$ on the negative-energy side. The reversal of conductance asymmetry and the flipping of noise asymmetry are not perfectly opposite, which may be attributed to the difference in tunnelling rates. But most importantly, in the zero-bias peak regime near the QPT, the noise asymmetry ($F^* \neq 1$) seen in this experiment reveals the trivial nature of the state.
	\clearpage
	
	\onecolumngrid
	
	\twocolumngrid
	
	\textbf{\textit{Discussion.---}} To confirm that our findings are genuine to the YSR states, we performed noise studies of zero-bias peaks across at least five different defect sites, with independent sample cleaves, distinct cooldown cycles, and different microtips, using both an oscilloscope and a spectrum analyser (see Supplemental Material \cite{SI}, Sec VII, for additional measurements). All these results show $F^* \neq 1$, ruling out contributions from tip condition, sample contamination, or instrumental artefacts. For defects \#A and \#B, the zero-bias peak  shows reduced noise ($F^* < 1$) on the left flank (negative bias) and enhanced noise ($F^* > 1$) on the right flank (positive bias), consistent with a superposition of two contributions, $G_{-}$ and $G_{+}$, with $G_{-} > G_{+}$. The opposite asymmetry ($G_{-} < G_{+}$) may in general also be expected. We show this in another impurity (defect \#C) for a zero-bias peak, as discussed in the Supplemental Material \cite{SI}, Sec VII, confirming that the noise asymmetry is intrinsic to the YSR state.
	
	Alternative effects, such as thermal noise or residual quasi-particle poisoning can also be ruled out. While local, energy-dependent heating could in principle increase the total noise, it cannot account for the observed noise suppression, which would require temperatures below our base temperature. Residual sub-gap quasiparticle states or tails of the superconducting coherence peaks may contribute to $dI/dV$ near zero-bias and to the noise. However, because the defect states we analysed are highly isolated in energy, this background is negligible in the relevant energy range. Even if the coherence peaks would contribute to the signal, their noise is Poissonian and would drive $F^*$ toward unity. Furthermore, other sub-gap states are separated from the peak(s) under study by a region of near-zero conductance, providing a sufficiently large energy window where the noise is exclusively carried by the YSR state.
	
	The question remains why the noise is so effective at revealing the particle-hole asymmetry of YSR states, even if the energy of the states is so close to zero as to be indistinguishable from it. By simple symmetry arguments, a YSR state exactly at zero should have $F^* = 1$. However, although a zero-bias peak is observed in the conductance spectroscopy, the peak is rarely truly located at zero-energy in the zero-temperature limit; due to finite thermal broadening, the feature appears at, or near zero-bias. Therefore, even though thermal mixing may occur, this slight offset from zero is sufficient for the noise to reveal $G_{+}\neq G_{-}$. The implication is significant: it demonstrates that unambiguous detection of impostor Majoranas is indeed possible through a single shot-noise measurement.
	
	\textbf{\textit{Conclusions.---}} Our work shows that shot-noise measurements are a powerful complement to standard spectroscopic probes. We demonstrate how deviations in the shot-noise immediately reveal that a sub-gap state is trivial, providing a robust test to identify impostor Majoranas. In doing so, we provide ample proof that conductance alone is not sufficient: a zero-bias peak could be (hidden) overlapping YSR pairs. In absence of a fully comprehensive set of measurements, the trivial nature of a (near-) zero-bias peak may be overlooked, whereas shot-noise reveals their true nature immediately with a single-shot measurement. Therefore, it is crucial to apply noise spectroscopy to other candidate Majorana platforms to (re)-assess whether the robust zero-bias peak, suggestive of a Majorana mode, is indeed genuine or instead fails the shot-noise test as we show here.
	\\
	\\
	\noindent \textit{Data Availability.---} Data for this work are available on reasonable request.
	\\
	\noindent \textit{Acknowledgements.---} We thank Véronique Brouet for her help with sample annealing. We also thank Marco Aprili, Andrej Mesaros, Simão Silva Cardoso and Pascal Simon for fruitful discussions. A.M. acknowledges funding from HORIZON-MSCA-2023-PF-01 (101152827). F.M. would like to acknowledge funding from the ANR (ANR-21-CE30-0017-01). The work at BNL was supported by the US Department of Energy, Office of Basic Energy Sciences, contract no. DOE-sc0012704.
	\\
	\noindent \textit{Author Contributions.---} A.M and F.M carried out the experiments, analysed the data, prepared the figures, and wrote the manuscript. G.D.G provided the samples. F.M supervised the project.

\end{document}